\documentclass[%
 reprint,
 amsmath,amssymb,
 aps,
]{revtex4-1}
\usepackage{graphicx}
\usepackage{color}
\usepackage{amsmath}
\usepackage{amssymb}
\usepackage{amsthm}
\usepackage{booktabs}
\usepackage{float}
\usepackage[dvipdfm,colorlinks,urlcolor=blue,linkcolor=blue,anchorcolor=blue,citecolor=blue]{hyperref}

\begin{document}

\title{Nonintegrability and thermalization of one-dimensional diatomic lattices}

\author{Weicheng Fu}
\author{Yong Zhang}
\email{yzhang75@xmu.edu.cn}
\author{Hong Zhao}
\email{zhaoh@xmu.edu.cn}

\affiliation{Department of Physics and Jiujiang Research Institute, Xiamen University, Xiamen 361005, Fujian, China}

\date{\today }
\begin{abstract}
       Nonintegrability is a necessary condition for the thermalization of a generic Hamiltonian system.~In practice, the integrability can be broken in various ways.~As illustrating examples, we numerically studied the thermalization behaviors of two types of one-dimensional (1D) diatomic chains in the thermodynamic limit.~One chain was the diatomic Toda chain whose nonintegrability was introduced by unequal masses.~The other chain was the diatomic Fermi-Pasta-Ulam-Tsingou-$\beta$ chain whose nonintegrability was introduced by quartic nonlinear interaction.~We found that these two different methods of destroying the integrability led to qualitatively different routes to thermalization, but the thermalization time, $T_{eq}$, followed the same law; $T_{eq}$ was inversely proportional to the square of the perturbation strength.~This law also agreed with the existing results of 1D monatomic lattices.~All these results imply that there is a universal law of thermalization that is independent of the method of breaking integrability.
\end{abstract}

\maketitle

\section{\label{sec:1}Introduction}

The famous ergodic hypothesis formulated by Boltzmann in the 1870s is at the foundation of statistical physics~\cite{1949Khinchin}.~In the 1950s, Fermi, in collaboration with Pasta, Ulam, and Tsingou (FPUT),  conducted the first numerical experiment to verify this hypothesis by observing the rate of mixing and thermalization in microscopic reversible mechanical systems~\cite{Fermi1955,dauxois:ensl-00202296}.~However, the result was contrary to general expectations.~The system far from equilibrium did not enter the expected thermalized state, but returned to nearly the initial nonequilibrium state.~Such a phenomenon is named the ``FPUT paradox''~\cite{dauxois:ensl-00202296}. This seminal work failed to observe the expected picture but it spurred many great mathematical and physical discoveries such as integrability~\cite{zakharov1991integrability}, soliton physics~\cite{PhysRevLett.15.240}, and deterministic chaos~\cite{doi:10.1063/1.1858115}.

More than half a century has passed; the literature on the subject is too varied to summarize here. The state of the art, updated to a few years ago, can be found in a collection of papers~\cite{Chaos2005}~and a status report~\cite{2008LNP728G}.~Nevertheless, if one tries to draw from these sources any clear conclusion about the mathematical status of the FPUT problem or the physical meaning of the results, one may remain rather confused~\cite{2008LNP728G}.~The literature is sometimes a bit confusing for various conditions due to the very rich dynamics of one-dimensional (1D) nonlinear chains.~For instance, based on Nekhoroshev's theory on exponential stability~\cite{Nekhoroshev_1977}, equipartition should always occur, and the thermalization time, $T_{eq}$, has a stretched exponential dependence on the energy density, $\varepsilon$; i.e., $T_{eq}\propto\exp(\varepsilon^{-a})$ with some positive $a$~\cite{fucito1982approach,PhysRevA.41.768,GALGANI1992334,BERCHIALLA2004167}.~However, some other researchers have shown that the thermalization time follows a power law; i.e., $T_{eq}\propto\varepsilon^{-a}$ with some positive $a$~\cite{PhysRevE.51.2877,*PhysRevE.54.2329,*PhysRevE.60.3781}.~Moreover, some results have shown a crossover from the stretched exponential law to a power law for various conditions~\cite{Benettin2011}.~Recently, the wave turbulence theory~\cite{1992kstbookZ,Majda1997,ZAKHAROV2001573,ZAKHAROV20041,2011LNP825N}~was applied to attack this problem~\cite{Onorato4208,PhysRevLett.120.144301,0295-5075-121-4-44003}.~It was shown analytically and numerically that the exact nontrivial multi-wave resonant interactions were responsible for the thermalization of short FPUT chains in the weakly nonlinear regime and this resulted in~$T_{eq}$~following a power law.~It was further conjectured that $T_{eq}$ still followed a power law with different exponents in the thermodynamic limit.

More recently, we have shown via extensive numerical calculations that the thermalization follows a universal power law in the thermodynamic limit~\cite{Our2018}.~This universal law, $T_{eq}\sim \gamma^{-2}$, applies generally to a class of 1D lattices with interaction potential $V(x)=x^2/2+\lambda x^n/n$, where $n\geq4$ is an integer and $\gamma= \lambda \varepsilon^{(n-2)/2}$ is the perturbation strength.~This scaling law also applies to another class of 1D lattices with symmetric interaction potential $V(x)=x^2/2+\lambda|x|^d/d$, where $d=m_1/m_2>2$ with $m_1$ and $m_2$ being two coprime integers and the perturbation strength $\gamma= \lambda \varepsilon^{(d-2)/2}$.~Furthermore, we numerically confirmed that this law of thermalization also held in the perturbed Toda lattices in the thermodynamic limit; i.e., $T_{eq}\propto\epsilon^{-2}$, where the perturbation strength, $\epsilon$, characterizes the distance between the perturbed potential and the Toda potential~\cite{Fu_2019}.~It has been shown that the key to identifying the universal exponent $-2$ is to select a suitable reference integrable system so that the perturbation strength can be defined accurately.~We noted that the thermalization in the Klein-Gordon lattice~\cite{0295-5075-121-4-44003,Pistone2018} also follows this law, though this lattice belongs to another class that possesses on-site potential.

In the above studies, the integrability of the system was broken by introducing nonlinearity.~However, there are various ways to destroy the integrability~\cite{toda1989theory}, including introducing impurities and raising dimensionality.~For example, the nonintegrability of a 1D diatomic Toda lattice was introduced by unequal masses~\cite{PhysRevA.12.1702}.~This is a somewhat natural perturbation in the sense that isotopic mass impurities do occur in  nature. There are many studies on the diatomic Toda chain~\cite{PhysRevA.23.959,PhysRevA.24.2826,Diederich_1985,aoki1995stationary,Mokross_1983,jackson1989thermal,PhysRevE.59.R1,PhysRevE.90.032134}, especially on its heat conduction behavior~\cite{Mokross_1983,jackson1989thermal,PhysRevE.59.R1,PhysRevE.90.032134}, but to our knowledge, the thermalization problem of diatomic Toda chains has not been discussed systematically.

In this study, we examined the thermalization rate of 1D diatomic Toda lattices in order to determine whether a universal law of $T_{eq}$ existed for this class of lattices as well, and if the answer was yes, how it differed from $T_{eq}\propto\epsilon^{-2}$ applicable to the 1D lattices in which the nonintegrability is introduced by nonlinearity.~For comparison, the diatomic FPUT-$\beta$ lattice was also studied, which is the diatomic harmonic lattice perturbed by the quartic nonlinearity; that is, its nonintegrability was introduced by nonlinearity, though it was a diatomic chain. In the following sections, we first introduce the models in the Sec.~\ref{sec:2}, then give the definitions of perturbation strength for the two models in Sec.~\ref{sec:3}.  The physical quantities and the numerical method are provided in Sec.~\ref{sec:4}. The numerical results are described and presented in Sec.~\ref{sec:5}, followed by the summary and discussions in Sec.~\ref{sec:6}.

\section{\label{sec:2}The models}

\begin{figure}[ht]
  \centering
  \includegraphics[width=1\columnwidth]{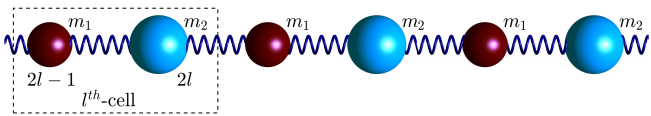}\\
  \caption{Structural diagram of the 1D diatomic chain.}\label{figDiatomicChain}
\end{figure}

We considered a 1D diatomic lattice consisting of $L$ unit cells, i.e., the total number of particles $N=2L$.~Each cell contained two particles of mass $m_1$ and $m_2$ ($m_1<m_2$) situated alternately at the position $2l-1$ and $2l$ in the $l$th unit cell. The physical configuration is illustrated in Fig.~\ref{figDiatomicChain}. Its Hamiltonian is
\begin{equation}\label{eqHariltonian}
  H=\sum_{j=1}^{N}\left[\frac{p_j^2}{2m_j}+V(q_j-q_{j-1})\right],
\end{equation}
where $p_j$, $m_j$, and $q_j$ are the momentum, mass, and displacement from the equilibrium position of the $j$th particle, respectively, and $V$ is the nearest-neighboring inter-particle interaction potential. Without losing generality, we set $m_1=1-\Delta m/2$, and $m_2=1+\Delta m/2$ ($|\Delta m|<2$ to guarantee positive masses) so that the mean value of the masses was fixed to be a unit; that is, the mass's density of the diatomic chain remained constant.

In our study, the Toda potential took the form of
\begin{equation}
   V_T(x)=\frac{e^{2 x}-2 x-1}{4}.
\end{equation}
The 1D diatomic Toda lattice increasingly approximates an integrable lattice as $\Delta m$ approaches zero.~Therefore, the diatomic Toda chain can be considered the perturbation of the Toda Hamiltonian.~In this study, we only focused on the thermalization problem in the near-integrable region; thus, $\Delta m\ll1$ was required.

For the sake of contrast, we also studied the diatomic FPUT-$\beta$ chain that possessed interaction potential
\begin{equation}
  V_\beta(x)=\frac{x^2}{2}+\frac{\beta  x^4}{4},
\end{equation}
where $\beta$ is a positive and free parameter.~The diatomic FPUT-$\beta$ chain is always nonintegrable even if $\Delta m=0$.~However, it will be the integrable diatomic harmonic lattice when $\beta=0$.~Naturally, the diatomic FPUT-$\beta$ chain should be regarded as the integrable diatomic harmonic lattice perturbed by the quartic nonlinearity.

\section{\label{sec:3}Definition of perturbation strength}
The Hamiltonian of a system can be written as
\begin{equation}
  H=H_0+H',
\end{equation}
where $H_0$ and $H'$ denote the integrable part and the perturbation, respectively.~Intuitively, the larger the perturbation is, the easier the system will be thermalized.~To accurately characterize the ability of the system to be thermalized, it was necessary to properly define the perturbation strength, that is, to select the appropriate $H_0$. From the Hamiltonian canonical equation, it is easy to prove that the dynamical system described by Eq.~(\ref{eqHariltonian}) is mathematically and strictly equivalent to the homogeneous chain with unit mass described by the following Hamiltonian
\begin{equation}\label{eqHariltonianR}
  H=\sum_{j=1}^{N}\left[\frac{v_j^2}{2}+\frac{1}{m_j}V(q_j-q_{j-1})\right],
\end{equation}
where $v_j$ is the velocity of $j$-th particle, and $1/m_j$ can be considered the renormalization coefficient of the force constant dependent on the lattice site.~Based on the Eq.~(\ref{eqHariltonianR}), the definitions of the perturbation strength for different cases are given below.

\subsection{The diatomic Toda chains}

For this case, the Hamiltonian of the Toda model was adopted as $H_0$, and
\begin{equation}\label{eqHT0}
  H_0=\sum_{j=1}^{N}\left[\frac{v_j^2}{2}+\frac{1}{m_2}V_T(q_j-q_{j-1})\right].
\end{equation}
Thus, we obtained the perturbation by comparing Eq.~(\ref{eqHariltonianR}) and Eq.~(\ref{eqHT0}) as
\begin{equation}
    H'=\frac{|m_2-m_1|}{m_1m_2}\sum_{j=1,3,5,\cdots}^{N}V_T(q_j-q_{j-1}),
\end{equation}
and the average strength of the perturbation for the fixed energy density
\begin{equation}
 \langle H'\rangle\propto\frac{|m_1-m_2|}{m_1m_2}=\frac{\Delta m}{1-\Delta m^2/4}\sim\Delta m,
\end{equation}
where $\langle \cdot \rangle$ denotes the ensemble average of the thermodynamic equilibrium state.~The perturbation strength being proportional to the mass difference illustrates the fact that the diatomic Toda chain is indeed the Toda Hamiltonian perturbed by unequal masses.

\subsection{The diatomic FPUT-$\beta$ chains}

For this case, the Hamiltonian of the diatomic harmonic model was adopted as $H_0$ and
\begin{equation}\label{eqHH0}
  H_0=\sum_{j=1}^{N}\left[\frac{v_j^2}{2}+\frac{1}{2m_j}(q_j-q_{j-1})^2\right].
\end{equation}
Comparing Eq.~(\ref{eqHariltonianR}) and Eq.~(\ref{eqHH0}), we obtained the perturbation
\begin{equation}
  H'=\beta\sum_{j=1}^{N}\frac{(q_j-q_{j-1})^4}{4m_j},
\end{equation}
and the average strength of the perturbation  for the fixed energy density
\begin{equation}
    \langle H'\rangle\propto\beta\frac{(m_1+m_2)}{m_1m_2}=\frac{\beta }{1-\Delta m^2/4}\sim\beta.
\end{equation}
The above expression clearly shows that the perturbation strength in the near-integrable region depended only on the nonlinear coefficients for this model.

\section{\label{sec:4}Physical quantities and numerical method}

\begin{figure}[t]
  \centering
  \includegraphics[width=1\columnwidth]{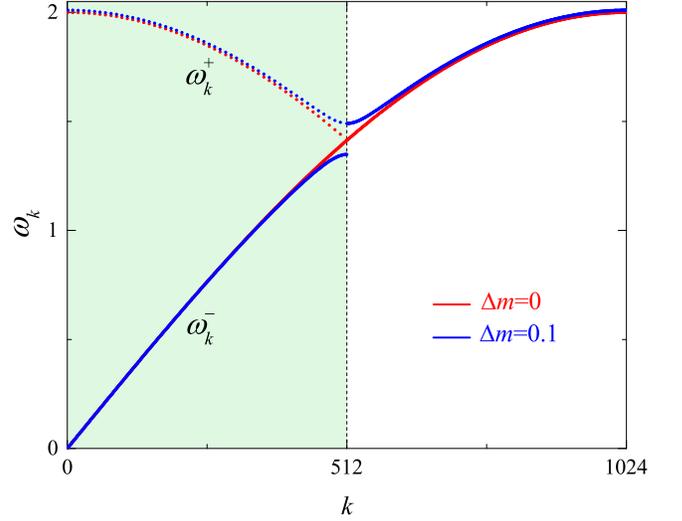}\\
  \caption{The dispersion relation of the diatomic chain with different $\Delta m$ values. The solid line and the dotted line, in the light green areas, correspond to the acoustic branch and the optical branch, respectively.~The unit cells' number $L=512$.~The solid lines at the right side of the vertical dashed line are the reflection of the optical branch.}
  \label{fig:omega}
\end{figure}

\begin{figure*}[t]
  \centering
  \includegraphics[width=2\columnwidth]{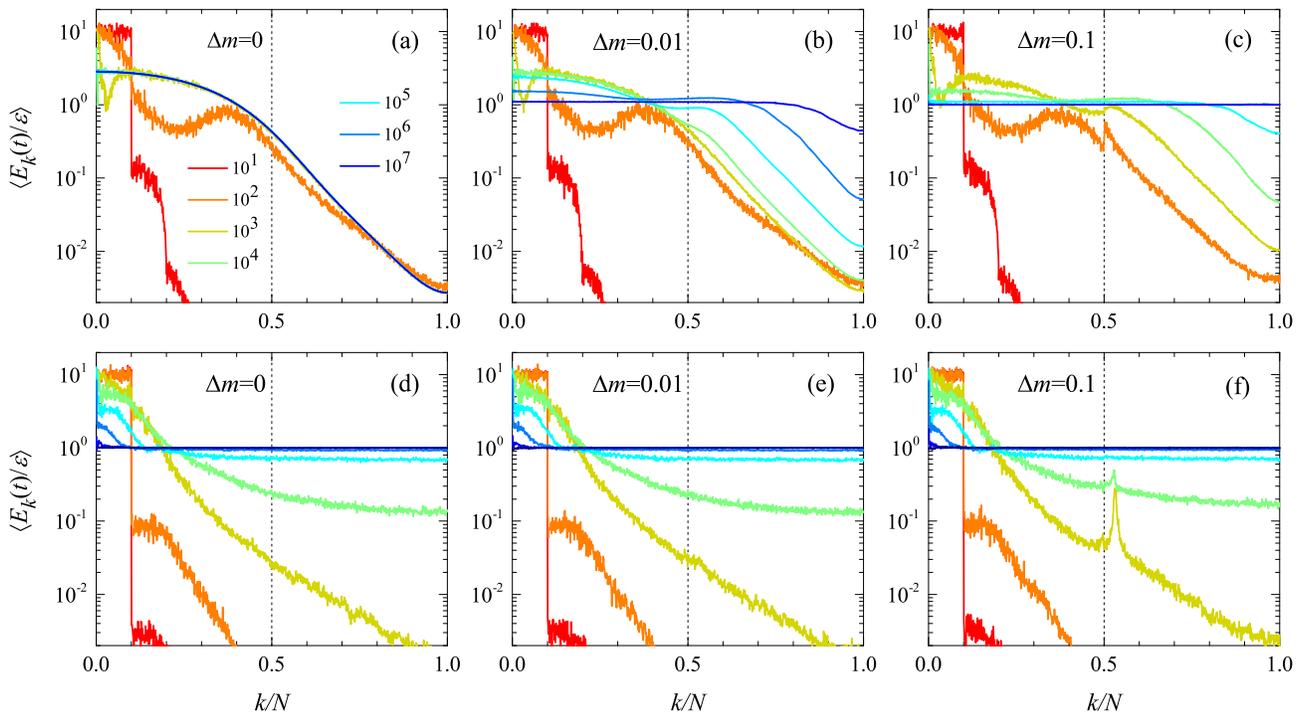}\\
  \caption{(a)-(c) The function~$\langle E_k(t)/\varepsilon\rangle$ versus $k/N$ at various times for the diatomic Toda chain with different mass perturbations, $\Delta m$; (d)-(f) show the results for the diatomic FPU-$\beta$ chain, with ensemble average measurements for $120$ different random choices of the phases. The total number of atoms~$N=1024$, and the energy density~$\varepsilon=0.01$ was fixed.}\label{fig:1}
\end{figure*}

The eigenvector, $\mathbf{u}_k$, of 1D diatomic chain for the fixed boundary conditions was derived in~\cite{Diatomic1974}. The element of $\mathbf{u}_k$ is
\begin{equation}\label{eq:Vec}
  u_{j,k}^{\pm}=\begin{cases}
    \dfrac{\sin\left(\frac{2lk\pi}{2L+1}\right)+\sin\left(\frac{2(l-1)k\pi}{2L+1}\right)}{\sin\left(\frac{2k\pi}{2L+1}\right)},&j=2l-1;\\
    \dfrac{\left[2-m_1\left(\omega_k^{\pm}\right)^2\right]\sin\left(\frac{2lk\pi}{2L+1}\right)}{\sin\left(\frac{2k\pi}{2L+1}\right)},&j=2l,
  \end{cases}
\end{equation}
where $\omega_k^{\pm}$ is the frequency of the $k$th eigenvector, as shown below:
\begin{equation}
\omega_k^{\pm}=\sqrt{\frac{m_1+m_2}{m_1m_2}\left[1\pm\sqrt{1-\frac{4m_1m_2}{(m_1+m_2)^2}\sin^2\left(\frac{k\pi}{2L+1}\right)}\right]}.
\end{equation}
In this formula, `$-$' and `$+$' correspond to the acoustic branch and the optical branch, respectively, and $k=1,2,\cdots,L$. It should be noted that the eigenvectors given in formula (\ref{eq:Vec}) correspond to two vectors of the same wave number $k$ that are not orthogonally normalized.~Therefore, by means of Gram-Schmidt's orthogonalization technique~\cite{leon2013gram}, the orthogonal and normalized eigenvector of the system can be obtained as follows:
\begin{align}
  \mathbf{U}_{k}^{+}&=\mathbf{u}_k^{+}/\|\mathbf{u}_k^{+}\|,\\
  \mathbf{U}_k^{-}&=\mathbf{u}_k^{-}-\frac{\langle \mathbf{u}_k^{-},\mathbf{U}_k^{+}\rangle}{\| \mathbf{U}_k^{+}\|^2}\mathbf{U}_k^{+},~\text{and}~\mathbf{U}_k^{-}=\frac{\mathbf{U}_k^{-}}{\|\mathbf{U}_k^{-}\|},
\end{align}
where $\|\cdot\|$ represents the length of a vector, and $\langle \mathbf{a},\mathbf{b}\rangle$ denotes the inner product of vectors $\mathbf{a}$ and $\mathbf{b}$. To produce a convenience of narration of the results, we sorted $N$ frequencies from small to large and removed the symbols `$\pm$'; i.e, $\omega_{k}=\omega_{k}^{-}$, and $\omega_{N-k+1}=\omega_{k}^{+}$ for $1\le k\le L$ (see Fig. \ref{fig:omega}).~The superscript `$\pm$' of the corresponding eigenvector was also removed.~Namely, $\mathbf{U}_{k}=\mathbf{U}_{k}^{-}$, and $\mathbf{U}_{N-k+1}=\mathbf{U}_{k}^{+}$. Thus, the normal modes of 1D diatomic lattice are defined as
\begin{align}
\begin{cases}
    Q_k&=\sum_{j=1}^{N}q_jU_{j,k},\\
    P_k&=\sum_{j=1}^{N}p_j/m_jU_{j,k},
\end{cases}\quad k=1,2,\cdots,N.
\end{align}
The energy of the $k$th normal mode is
\begin{equation}
    E_k=\frac{1}{2}\left(P_k^2+\omega_k^2Q_k^2\right),
\end{equation}
and a phase $\varphi_k$ is defined via
\begin{equation}
   Q_k=\sqrt{2E_k/\omega_k^2}\sin{\left(\varphi_k\right)},~P_k=\sqrt{2E_k}\cos{\left(\varphi_k\right)}.
\end{equation}
Following the definition of equipartition, one expects
\begin{equation}
  \lim_{T\rightarrow\infty}\bar{E}_k(T)\simeq\varepsilon, \quad k=1,~\cdots,~N,
\end{equation}
where $\varepsilon$ is the energy per particle and $\bar{E}_k(T)$ represents the time average of $E_k$ up to time $T$; i.e.,
\begin{equation}\label{eq:EkT}
  \bar{E}_k(T)=\frac{1}{(1-\mu)T}\int_{\mu T}^TE_k(P(t),Q(t))dt.
\end{equation}
In the formula above, $\mu\in[0,1)$ controls the size of the time average window. In our numerical simulations, $\mu=2/3$ was fixed, which could not only speed up the calculations, but also had the advantage of a quicker loss of the memory of the very special initial state, as proposed in~\cite{Benettin2011}.

Based on the defined $\bar{E}_k(T)$, we needed to introduce a parameter to measure how close the system was to equipartition. A parameter frequently used for this purpose is the effective relative number of degrees of freedom~\cite{PhysRevA.31.1039, GOEDDE1992200}. We employed the quantity $\xi(t)$, as described in Ref.~\cite{Benettin2011}, i.e.,
\begin{equation}\label{eq:xi}
  \xi(t)=\tilde{\xi}(t)\frac{e^{\eta(t)}}{L},\quad \eta(t)=-\sum_{k>L}^{N}w_k(t)\log[w_k(t)]
\end{equation}
and
\begin{equation}
  \tilde{\xi}(t)=\frac{\sum_{k>L}^{N}\bar{E}_k(t)}{\frac{1}{2}\sum_{1\leq{k}\leq{N}}\bar{E}_k(t)},~
  w_k(t)=\frac{\bar{E}_k(t)}{\sum_{j>L}^{N}\bar{E}_j(t)}.
\end{equation}
When equipartition was approached, $\xi$ will saturated at $1$.

To integrate the motion equations numerically, we used the eighth-order Yoshida method~\cite{YOSHIDA1990262}. The typical time step was $\Delta t=0.1$. To suppress fluctuations, the ensemble average was done over phases uniformly distributed in $[0,2\pi]$. In all our calculations below, the lowest $10\%$ of the frequency modes were excited. We have checked and verified that no qualitative difference will be resulted in when the percentage of the excited modes was changed.

\section{\label{sec:5}Numerical results}

In Fig.~\ref{fig:1}(a), the results of $\langle E_k(t)/\varepsilon\rangle$ versus $k/N$ for the Toda chain are presented. It can be seen that only a small portion of the energy spread quickly from the initially excited low-frequency modes to the high-frequency modes. After this, the energy profile kept a stable form.~This suggests that for the Toda model, the thermalized state can never be reached~\cite{Fu_2019}.~In Fig.~\ref{fig:1}(b), the results for the diatomic Toda chain with $\Delta m=0.01$ are plotted. It can be seen that the energy of the low-frequency modes (acoustic modes) was shared quickly while the energy of the high-frequency modes (optical modes) increased very slowly.~Namely, the acoustic modes entered the equipartition state first, while the optical modes entered the equipartition state last.~The system was eventually thermalized.~Figure~\ref{fig:1}(c) shows the results of the diatomic Toda chain with $\Delta m=0.1$.~It can be seen that the system reached the thermalized state faster.~The system was fully thermalized at time $T\sim 10^6$, when $\langle E_k/\varepsilon\rangle=1$. As a comparison, Figs.~\ref{fig:1}(d)-(f) show the results of the diatomic FPUT-$\beta$ chain with $\Delta m=0$, $0.01$, and $0.1$, respectively. It can be seen that the results had a tiny difference, and all the systems were fully thermalized at nearly the same time, e.g., $T\sim 10^8$, when $\langle E_k/\varepsilon\rangle=1$. However, these results differ qualitatively from those of the diatomic Toda chains.~In the FPUT-$\beta$ model, the high frequency modes (optical modes) entered the equipartition state first, while the low frequency modes (acoustic modes) entered the equipartition state last.~This phenomenon is also observed in the 1D monatomic chains perturbed by nonlinearity~\cite{Benettin2011,Fu_2019}.~We judged that this qualitative difference in the thermalization process was caused by the difference between the mass perturbation and the nonlinearity perturbation.

\begin{figure}[t]
  \centering
  \includegraphics[width=1\columnwidth]{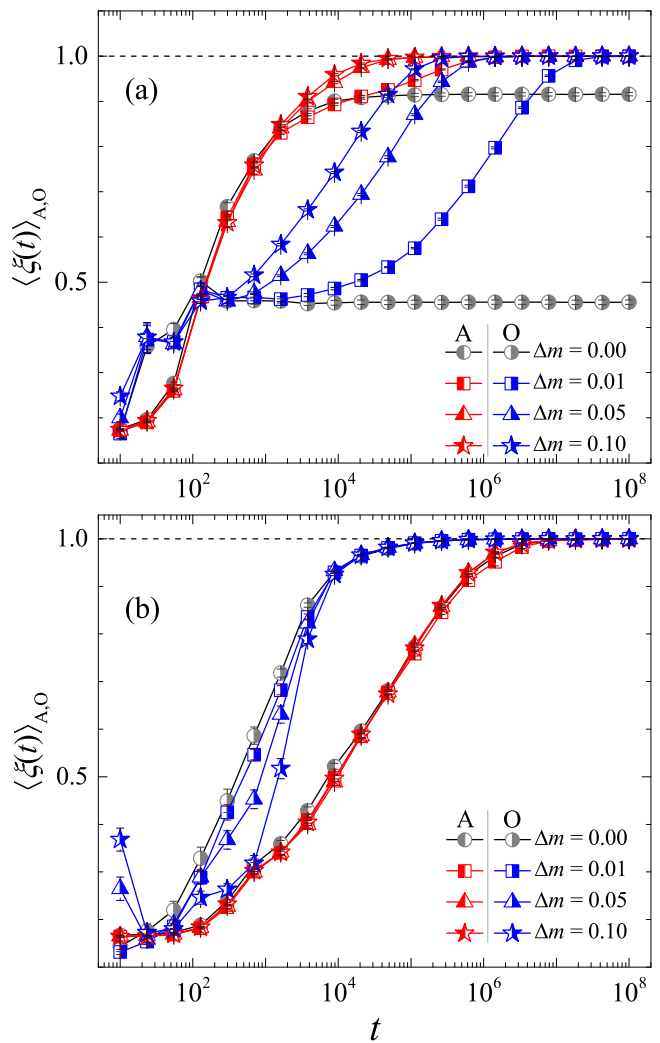}\\
  \caption{(a) The function of $\langle\xi(t)\rangle_{A,O}$ for the diatomic Toda chain with different mass perturbations on a semi-log scale.~The letters A and O in the legend indicate the results of the acoustic/optical modes.~The gray circles represent the results of the monatomic chain, with ensemble average measurements of $24$ different random choices of the phases.~The total number of atoms~$N=1024$, and the energy density~$\varepsilon=0.01$ was fixed.~(b) The results of the diatomic FPUT-$\beta$ chains.}\label{fig:2A}
\end{figure}

In order to further clarify the difference of the rate of the energy sharing between the acoustic modes and the optical modes, we defined the quantity $\xi(t)_{A,O}$ for the acoustic modes and the optical modes as
\begin{equation}
  \xi(t)_{A}=\frac{e^{\eta(t)}}{L},\quad \eta(t)=-\sum_{k=1}^{L}w_k(t)\log[w_k(t)],
\end{equation}
where $w_k(t)=\frac{\bar{E}_k(t)}{\sum_{j=1}^{L}\bar{E}_j(t)}$, and
\begin{equation}
  \xi(t)_{O}=\frac{e^{\eta(t)}}{L},\quad \eta(t)=-\sum_{k>L}^{N}w_k(t)\log[w_k(t)],
\end{equation}
where $w_k(t)=\frac{\bar{E}_k(t)}{\sum_{j>L}^{N}\bar{E}_j(t)}$. Figure~\ref{fig:2A}(a) shows the results for the diatomic Toda chains with fixed energy density $\varepsilon=0.01$, and different $\Delta m$; and~Fig.~\ref{fig:2A}(b) shows the results for the diatomic FPUT-$\beta$ chains.~It can be seen that $\xi(t)_{A,O}$ quickly reached a constant value less than $1$ for the monatomic Toda chain; that is, it never reached an equipartition state due to its integrability. However, the acoustic and optical modes of the FPUT-$\beta$ model would eventually enter the state of energy sharing. Except for this difference, it can be clearly seen that there was a qualitative difference in the route to thermalization for these two types of diatomic chains. For example, in the diatomic Toda chain, the energy of the acoustic modes entered the equipartition state first, then the energy of the optical modes entered the equipartition state. In contrast, the thermalization process of the FPUT-$\beta$ chain had the opposite order. In the following sections, we will further examine the systematic study of the relationship between the thermalization time and the perturbation strength of the system.

\begin{figure}[t]
  \centering
  \includegraphics[width=1\columnwidth]{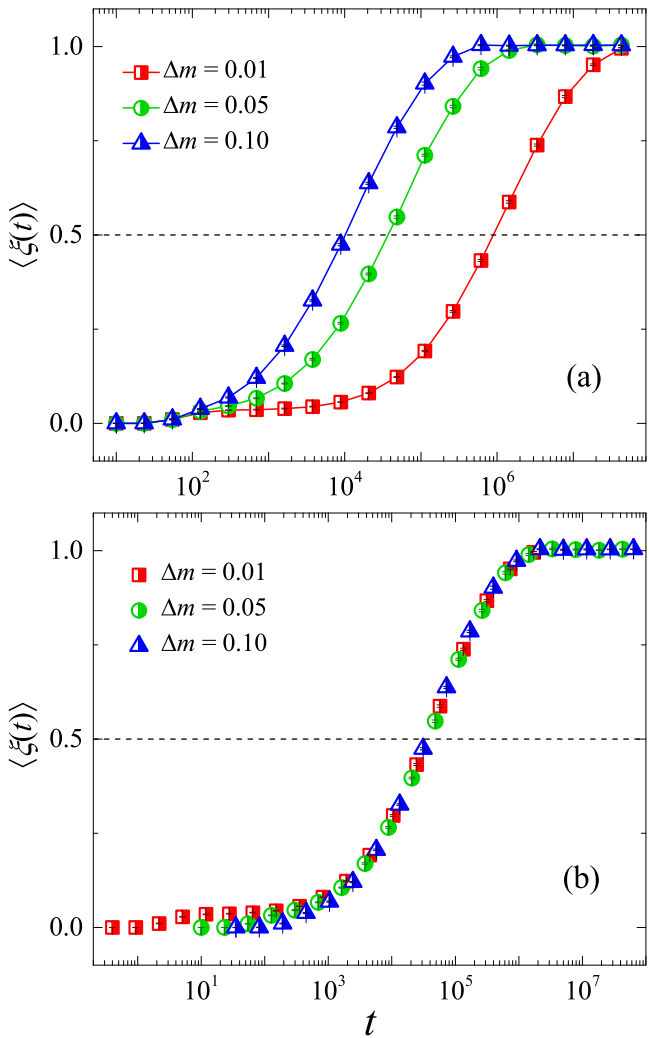}\\
  \caption{(a) The function of $\langle\xi(t)\rangle$ for the diatomic Toda chain with different mass perturbations on the semi-log scale and ensemble average measurements for $24$ different random choices of the phases. The total number of atoms~$N=1024$, and the energy density~$\varepsilon=0.01$ was fixed.~(b) The same as (a) but the curves are shifted properly in the horizontal direction (with that for $\Delta m=0.05$ unshifted) so that they perfectly overlap with each other.}\label{fig:2}
\end{figure}

To obtain the thermalization time, we studied the properties of $\langle \xi(t)\rangle$ defined by Eq.~(\ref{eq:xi}).~Figure~\ref{fig:2}(a) shows the results for the diatomic Toda chain with the fixed energy density $\varepsilon=0.01$, and different $\Delta m$. It should be noted that on a sufficiently large time scale, all values of $\langle \xi(t)\rangle$ increased from $0$ to $1$ with very similar sigmoidal profiles. This suggests that energy equipartition was finally achieved. Additionally, when $\Delta m$ decreased, the time required to reach the thermalized state increased. We adopt the definition of the equipartition time, $T_{eq}$, as the time when $\langle \xi(t)\rangle$ reached the threshold value $0.5$, as described in~\cite{Benettin2011}. By assuming that the threshold value $0.5$ was artificial, it did not influence the scaling law of $T_{eq}$. This can be seen from Fig.~\ref{fig:2}(b), where the sigmoidal profiles in Fig.~\ref{fig:2}(a) overlap with each other upon suitable shifts, which suggests that the concrete threshold value did not affect the scaling exponent of $T_{eq}$. With these preparations, we were ready to present the results of $T_{eq}$.

\begin{figure}[t]
  \centering
  \includegraphics[width=1\columnwidth]{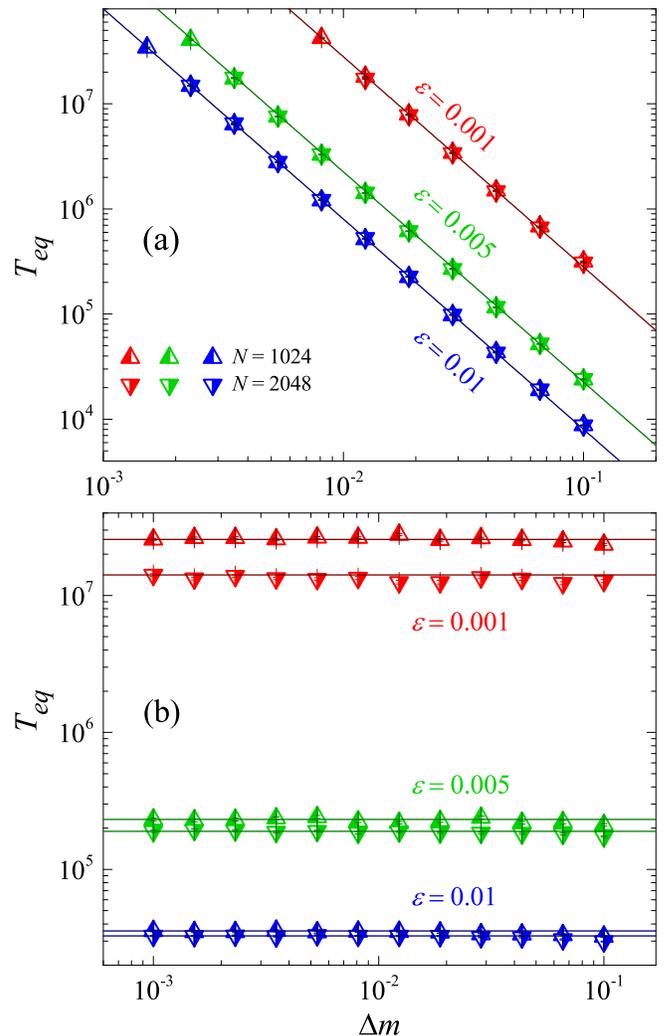}\\
  \caption{(a) The equipartition time $T_{eq}$ of the diatomic Toda chain as a function of $\Delta m$ for different energy densities and system sizes for a log-log scale.~The solid lines with slope $-2$ are drawn for reference.~(b) shows the results of the diatomic FPUT-$\beta$ chain with the same conditions in (a).~There were ensemble average measurements on $24$ different random choices of the phases.}\label{fig:3a}
\end{figure}

In Fig.~\ref{fig:3a}(a), the numerical results of $T_{eq}$ as a function of $\Delta m$ are shown in log-log scale for the diatomic Toda chain with different energy densities $\varepsilon=0.001$ (red triangles), $\varepsilon=0.005$ (green triangles), and $\varepsilon=0.01$ (blue triangles); system size $N=1024$ (up triangles), and $N=2048$ (down triangles). It should be noted that all the points fall on the lines with a slope of $-2$, suggesting $T_{eq}\propto\Delta m^{-2}$; i.e., $T_{eq}$ was inversely proportional to the square of perturbation strength. In addition, the results of different sizes coincided completely, i.e., the size effect disappeared in our calculated range. For contrast, the numerical results of the diatomic FPUT-$\beta$ chain with the same conditions are presented in Fig.~\ref{fig:3a}(b). It can be clearly seen that all the points fall on the lines with a slope of $0$, suggesting $T_{eq}\propto\Delta m^{0}$; i.e., $T_{eq}$ did not depend on $\Delta m$ since the nonintegrability of the diatomic FPUT-$\beta$ chain hardly changed with the variation of $\Delta m$. This means that the results of the diatomic FPUT-$\beta$ chains are completely identical to those of the monatomic FPUT-$\beta$ chains; i.e., $T_{eq}\propto\beta^{-2}$~\cite{Our2018,Pistone2018}. In addition, it should be noted that the results vary with the size of system under the same conditions, and the lower the energy density was, the more obvious the difference was; i.e., the stronger the size effect was. The phenomenon in which the symmetric model has a stronger size effect than the asymmetric model has been observed in the literatures~\cite{Benettin2011,Our2018}.

\section{\label{sec:6}Summary and discussions}

In this study, we examined the thermalization behaviors of two types of 1D diatomic chains; i.e., the mass perturbed Toda chain (the diatomic Toda chain) and the quartic nonlinearity perturbed diatomic harmonic chain (the diatomic FPUT-$\beta$ chain).~We showed that the acoustic mode in the diatomic Toda chain entered the state of energy sharing first, and the optical mode entered the equipartition state later.~In contrast, for the diatomic FPUT-$\beta$ chain, the optical mode entered the state of energy sharing first, while the acoustic mode entered the equipartition state later.~We considered that the qualitative difference in the thermalization process originated from the different methods of perturbation.~Although these qualitative differences existed, the scaling law of thermalization time followed the same rule; the thermalization time was inversely proportional to the square of the perturbation strength.~This law was also consistent with the thermalization law of 1D monatomic chains with nonlinearity perturbations~\cite{Our2018,Pistone2018,Fu_2019}.~All these results agree with the fact that the thermalization of a 1D chain is universal, and this universality is independent of the way in which nonintegrability is introduced.~Specifically, the key to identifying the universal law is the selection of a suitable $H_0$ as the reference integrable system, so that the perturbation strength that reflects the ability of the system to be thermalized is defined accurately.

\section*{Acknowledgment}

This work is supported by NSFC (Grant No. 11335006).

\section*{References}

\bibliography{DiatomicReference}

\begin{thebibliography}{43}%
\makeatletter
\providecommand \@ifxundefined [1]{%
 \@ifx{#1\undefined}
}%
\providecommand \@ifnum [1]{%
 \ifnum #1\expandafter \@firstoftwo
 \else \expandafter \@secondoftwo
 \fi
}%
\providecommand \@ifx [1]{%
 \ifx #1\expandafter \@firstoftwo
 \else \expandafter \@secondoftwo
 \fi
}%
\providecommand \natexlab [1]{#1}%
\providecommand \enquote  [1]{``#1''}%
\providecommand \bibnamefont  [1]{#1}%
\providecommand \bibfnamefont [1]{#1}%
\providecommand \citenamefont [1]{#1}%
\providecommand \href@noop [0]{\@secondoftwo}%
\providecommand \href [0]{\begingroup \@sanitize@url \@href}%
\providecommand \@href[1]{\@@startlink{#1}\@@href}%
\providecommand \@@href[1]{\endgroup#1\@@endlink}%
\providecommand \@sanitize@url [0]{\catcode `\\12\catcode `\$12\catcode
  `\&12\catcode `\#12\catcode `\^12\catcode `\_12\catcode `\%12\relax}%
\providecommand \@@startlink[1]{}%
\providecommand \@@endlink[0]{}%
\providecommand \url  [0]{\begingroup\@sanitize@url \@url }%
\providecommand \@url [1]{\endgroup\@href {#1}{\urlprefix }}%
\providecommand \urlprefix  [0]{URL }%
\providecommand \Eprint [0]{\href }%
\providecommand \doibase [0]{http://dx.doi.org/}%
\providecommand \selectlanguage [0]{\@gobble}%
\providecommand \bibinfo  [0]{\@secondoftwo}%
\providecommand \bibfield  [0]{\@secondoftwo}%
\providecommand \translation [1]{[#1]}%
\providecommand \BibitemOpen [0]{}%
\providecommand \bibitemStop [0]{}%
\providecommand \bibitemNoStop [0]{.\EOS\space}%
\providecommand \EOS [0]{\spacefactor3000\relax}%
\providecommand \BibitemShut  [1]{\csname bibitem#1\endcsname}%
\let\auto@bib@innerbib\@empty
%</preamble>
\bibitem [{\citenamefont {Khinchin}(1949)}]{1949Khinchin}%
  \BibitemOpen
  \bibfield  {author} {\bibinfo {author} {\bibfnamefont {A.~I.}\ \bibnamefont
  {Khinchin}},\ }\href@noop {} {\emph {\bibinfo {title} {Mathematical
  Foundations of Statistical Mechanics}}}\ (\bibinfo  {publisher} {Dover, New
  York},\ \bibinfo {year} {1949})\BibitemShut {NoStop}%
\bibitem [{\citenamefont {Fermi}\ \emph {et~al.}(1955)\citenamefont {Fermi},
  \citenamefont {Pasta},\ and\ \citenamefont {Ulam}}]{Fermi1955}%
  \BibitemOpen
  \bibfield  {author} {\bibinfo {author} {\bibfnamefont {E.}~\bibnamefont
  {Fermi}}, \bibinfo {author} {\bibfnamefont {P.}~\bibnamefont {Pasta}}, \ and\
  \bibinfo {author} {\bibfnamefont {S.}~\bibnamefont {Ulam}},\ }\href@noop {}
  {\bibfield  {journal} {\bibinfo  {journal} {Los Alamos Scientific Laboratory,
  Report No. LA-1940}\ } (\bibinfo {year} {1955})}\BibitemShut {NoStop}%
\bibitem [{\citenamefont {Dauxois}(2008)}]{dauxois:ensl-00202296}%
  \BibitemOpen
  \bibfield  {author} {\bibinfo {author} {\bibfnamefont {T.}~\bibnamefont
  {Dauxois}},\ }\href {\doibase 10.1063/1.2835154} {\bibfield  {journal}
  {\bibinfo  {journal} {{Phys. Today}}\ ,\ \bibinfo {pages} {55}} (\bibinfo
  {year} {2008})}\BibitemShut {NoStop}%
\bibitem [{\citenamefont {Zakharov}(1991)}]{zakharov1991integrability}%
  \BibitemOpen
  \bibfield  {author} {\bibinfo {author} {\bibfnamefont {V.~E.}\ \bibnamefont
  {Zakharov}},\ }\href@noop {} {\emph {\bibinfo {title} {What is
  integrability?}}}\ (\bibinfo  {publisher} {Springer,Berlin},\ \bibinfo {year}
  {1991})\BibitemShut {NoStop}%
\bibitem [{\citenamefont {Zabusky}\ and\ \citenamefont
  {Kruskal}(1965)}]{PhysRevLett.15.240}%
  \BibitemOpen
  \bibfield  {author} {\bibinfo {author} {\bibfnamefont {N.~J.}\ \bibnamefont
  {Zabusky}}\ and\ \bibinfo {author} {\bibfnamefont {M.~D.}\ \bibnamefont
  {Kruskal}},\ }\href {\doibase 10.1103/PhysRevLett.15.240} {\bibfield
  {journal} {\bibinfo  {journal} {Phys. Rev. Lett.}\ }\textbf {\bibinfo
  {volume} {15}},\ \bibinfo {pages} {240} (\bibinfo {year} {1965})}\BibitemShut
  {NoStop}%
\bibitem [{\citenamefont {Zaslavsky}(2005)}]{doi:10.1063/1.1858115}%
  \BibitemOpen
  \bibfield  {author} {\bibinfo {author} {\bibfnamefont {G.~M.}\ \bibnamefont
  {Zaslavsky}},\ }\href {\doibase 10.1063/1.1858115} {\bibfield  {journal}
  {\bibinfo  {journal} {Chaos}\ }\textbf {\bibinfo {volume} {15}},\ \bibinfo
  {pages} {015103} (\bibinfo {year} {2005})}\BibitemShut {NoStop}%
\bibitem [{Cha(2005)}]{Chaos2005}%
  \BibitemOpen
  \href@noop {} {\bibfield  {journal} {\bibinfo  {journal} {Chaos focus issue:
  The ``Fermi-Pasta-Ulam'' problem---the first 50 years,~Chaos}\ }\textbf
  {\bibinfo {volume} {15}} (\bibinfo {year} {2005})}\BibitemShut {NoStop}%
\bibitem [{\citenamefont {{Gallavotti}}(2008)}]{2008LNP728G}%
  \BibitemOpen
  \bibinfo {editor} {\bibfnamefont {G.}~\bibnamefont {{Gallavotti}}},\ ed.,\
  \href {\doibase 10.1007/978-3-540-72995-2} {\emph {\bibinfo {title} {The
  Fermi-Pasta-Ulam Problem Lecture Notes in Physics}}},\ \bibinfo {series}
  {Berlin Springer Verlag}, Vol.\ \bibinfo {volume} {728}\ (\bibinfo {year}
  {2008})\BibitemShut {NoStop}%
\bibitem [{\citenamefont {Nekhoroshev}(1977)}]{Nekhoroshev_1977}%
  \BibitemOpen
  \bibfield  {author} {\bibinfo {author} {\bibfnamefont {N.~N.}\ \bibnamefont
  {Nekhoroshev}},\ }\href {\doibase 10.1070/rm1977v032n06abeh003859} {\bibfield
   {journal} {\bibinfo  {journal} {Russ. Math. Surv.}\ }\textbf {\bibinfo
  {volume} {32}},\ \bibinfo {pages} {1} (\bibinfo {year} {1977})}\BibitemShut
  {NoStop}%
\bibitem [{\citenamefont {Fucito}\ \emph {et~al.}(1982)\citenamefont {Fucito},
  \citenamefont {Marchesoni}, \citenamefont {Marinari}, \citenamefont {Parisi},
  \citenamefont {Peliti}, \citenamefont {Ruffo},\ and\ \citenamefont
  {Vulpiani}}]{fucito1982approach}%
  \BibitemOpen
  \bibfield  {author} {\bibinfo {author} {\bibfnamefont {F.}~\bibnamefont
  {Fucito}}, \bibinfo {author} {\bibfnamefont {F.}~\bibnamefont {Marchesoni}},
  \bibinfo {author} {\bibfnamefont {E.}~\bibnamefont {Marinari}}, \bibinfo
  {author} {\bibfnamefont {G.}~\bibnamefont {Parisi}}, \bibinfo {author}
  {\bibfnamefont {L.}~\bibnamefont {Peliti}}, \bibinfo {author} {\bibfnamefont
  {S.}~\bibnamefont {Ruffo}}, \ and\ \bibinfo {author} {\bibfnamefont
  {A.}~\bibnamefont {Vulpiani}},\ }\href {\doibase
  10.1051/jphys:01982004305070700} {\bibfield  {journal} {\bibinfo  {journal}
  {J. Phys.}\ }\textbf {\bibinfo {volume} {43}},\ \bibinfo {pages} {707}
  (\bibinfo {year} {1982})}\BibitemShut {NoStop}%
\bibitem [{\citenamefont {Pettini}\ and\ \citenamefont
  {Landolfi}(1990)}]{PhysRevA.41.768}%
  \BibitemOpen
  \bibfield  {author} {\bibinfo {author} {\bibfnamefont {M.}~\bibnamefont
  {Pettini}}\ and\ \bibinfo {author} {\bibfnamefont {M.}~\bibnamefont
  {Landolfi}},\ }\href {\doibase 10.1103/PhysRevA.41.768} {\bibfield  {journal}
  {\bibinfo  {journal} {Phys. Rev. A}\ }\textbf {\bibinfo {volume} {41}},\
  \bibinfo {pages} {768} (\bibinfo {year} {1990})}\BibitemShut {NoStop}%
\bibitem [{\citenamefont {Galgani}\ \emph {et~al.}(1992)\citenamefont
  {Galgani}, \citenamefont {Giorgilli}, \citenamefont {Martinoli},\ and\
  \citenamefont {Vanzini}}]{GALGANI1992334}%
  \BibitemOpen
  \bibfield  {author} {\bibinfo {author} {\bibfnamefont {L.}~\bibnamefont
  {Galgani}}, \bibinfo {author} {\bibfnamefont {A.}~\bibnamefont {Giorgilli}},
  \bibinfo {author} {\bibfnamefont {A.}~\bibnamefont {Martinoli}}, \ and\
  \bibinfo {author} {\bibfnamefont {S.}~\bibnamefont {Vanzini}},\ }\href
  {\doibase https://doi.org/10.1016/0167-2789(92)90074-W} {\bibfield  {journal}
  {\bibinfo  {journal} {Physica D}\ }\textbf {\bibinfo {volume} {59}},\
  \bibinfo {pages} {334 } (\bibinfo {year} {1992})}\BibitemShut {NoStop}%
\bibitem [{\citenamefont {Berchialla}\ \emph {et~al.}(2004)\citenamefont
  {Berchialla}, \citenamefont {Giorgilli},\ and\ \citenamefont
  {Paleari}}]{BERCHIALLA2004167}%
  \BibitemOpen
  \bibfield  {author} {\bibinfo {author} {\bibfnamefont {L.}~\bibnamefont
  {Berchialla}}, \bibinfo {author} {\bibfnamefont {A.}~\bibnamefont
  {Giorgilli}}, \ and\ \bibinfo {author} {\bibfnamefont {S.}~\bibnamefont
  {Paleari}},\ }\href {\doibase 10.1016/j.physleta.2003.11.052} {\bibfield
  {journal} {\bibinfo  {journal} {Phys. Lett. A}\ }\textbf {\bibinfo {volume}
  {321}},\ \bibinfo {pages} {167 } (\bibinfo {year} {2004})}\BibitemShut
  {NoStop}%
\bibitem [{\citenamefont {DeLuca}\ \emph {et~al.}(1995)\citenamefont {DeLuca},
  \citenamefont {Lichtenberg},\ and\ \citenamefont {Ruffo}}]{PhysRevE.51.2877}%
  \BibitemOpen
  \bibfield  {author} {\bibinfo {author} {\bibfnamefont {J.}~\bibnamefont
  {DeLuca}}, \bibinfo {author} {\bibfnamefont {A.~J.}\ \bibnamefont
  {Lichtenberg}}, \ and\ \bibinfo {author} {\bibfnamefont {S.}~\bibnamefont
  {Ruffo}},\ }\href {\doibase 10.1103/PhysRevE.51.2877} {\bibfield  {journal}
  {\bibinfo  {journal} {Phys. Rev. E}\ }\textbf {\bibinfo {volume} {51}},\
  \bibinfo {pages} {2877} (\bibinfo {year} {1995})}\BibitemShut {NoStop}%
\bibitem [{\citenamefont {\textit{ibid}}(1996)}]{PhysRevE.54.2329}%
  \BibitemOpen
  \bibfield  {author} {\bibinfo {author} {\bibnamefont {\textit{ibid}}},\
  }\href {\doibase 10.1103/PhysRevE.54.2329} {\ \textbf {\bibinfo {volume}
  {54}},\ \bibinfo {pages} {2329} (\bibinfo {year} {1996})}\BibitemShut
  {NoStop}%
\bibitem [{\citenamefont {\textit{ibid}}(1999)}]{PhysRevE.60.3781}%
  \BibitemOpen
  \bibfield  {author} {\bibinfo {author} {\bibnamefont {\textit{ibid}}},\
  }\href {\doibase 10.1103/PhysRevE.60.3781} {\ \textbf {\bibinfo {volume}
  {60}},\ \bibinfo {pages} {3781} (\bibinfo {year} {1999})}\BibitemShut
  {NoStop}%
\bibitem [{\citenamefont {Benettin}\ and\ \citenamefont
  {Ponno}(2011)}]{Benettin2011}%
  \BibitemOpen
  \bibfield  {author} {\bibinfo {author} {\bibfnamefont {G.}~\bibnamefont
  {Benettin}}\ and\ \bibinfo {author} {\bibfnamefont {A.}~\bibnamefont
  {Ponno}},\ }\href {\doibase 10.1007/s10955-011-0277-9} {\bibfield  {journal}
  {\bibinfo  {journal} {J. Stat. Phys.}\ }\textbf {\bibinfo {volume} {144}},\
  \bibinfo {pages} {793} (\bibinfo {year} {2011})}\BibitemShut {NoStop}%
\bibitem [{\citenamefont {{Zakharov}}\ \emph {et~al.}(1992)\citenamefont
  {{Zakharov}}, \citenamefont {{L'Vov}},\ and\ \citenamefont
  {{Falkovich}}}]{1992kstbookZ}%
  \BibitemOpen
  \bibfield  {author} {\bibinfo {author} {\bibfnamefont {V.~E.}\ \bibnamefont
  {{Zakharov}}}, \bibinfo {author} {\bibfnamefont {V.~S.}\ \bibnamefont
  {{L'Vov}}}, \ and\ \bibinfo {author} {\bibfnamefont {G.}~\bibnamefont
  {{Falkovich}}},\ }\href {\doibase 10.1007/978-3-642-50052-7} {\emph {\bibinfo
  {title} {Kolmogorov spectra of turbulence I.~Wave turbulence.}}},\ Springer,
  Berlin (Germany),\ (\bibinfo {year} {1992})\BibitemShut {NoStop}%
\bibitem [{\citenamefont {Majda}\ \emph {et~al.}(1997)\citenamefont {Majda},
  \citenamefont {McLaughlin},\ and\ \citenamefont {Tabak}}]{Majda1997}%
  \BibitemOpen
  \bibfield  {author} {\bibinfo {author} {\bibfnamefont {A.~J.}\ \bibnamefont
  {Majda}}, \bibinfo {author} {\bibfnamefont {D.~W.}\ \bibnamefont
  {McLaughlin}}, \ and\ \bibinfo {author} {\bibfnamefont {E.~G.}\ \bibnamefont
  {Tabak}},\ }\href {\doibase 10.1007/BF02679124} {\bibfield  {journal}
  {\bibinfo  {journal} {J. Nonlinear Sci.}\ }\textbf {\bibinfo {volume} {7}},\
  \bibinfo {pages} {9} (\bibinfo {year} {1997})}\BibitemShut {NoStop}%
\bibitem [{\citenamefont {Zakharov}\ \emph {et~al.}(2001)\citenamefont
  {Zakharov}, \citenamefont {Guyenne}, \citenamefont {Pushkarev},\ and\
  \citenamefont {Dias}}]{ZAKHAROV2001573}%
  \BibitemOpen
  \bibfield  {author} {\bibinfo {author} {\bibfnamefont {V.}~\bibnamefont
  {Zakharov}}, \bibinfo {author} {\bibfnamefont {P.}~\bibnamefont {Guyenne}},
  \bibinfo {author} {\bibfnamefont {A.}~\bibnamefont {Pushkarev}}, \ and\
  \bibinfo {author} {\bibfnamefont {F.}~\bibnamefont {Dias}},\ }\href {\doibase
  10.1016/S0167-2789(01)00194-4} {\bibfield  {journal} {\bibinfo  {journal}
  {Physica D}\ }\textbf {\bibinfo {volume} {152-153}},\ \bibinfo {pages} {573 }
  (\bibinfo {year} {2001})}\BibitemShut {NoStop}%
\bibitem [{\citenamefont {Zakharov}\ \emph {et~al.}(2004)\citenamefont
  {Zakharov}, \citenamefont {Dias},\ and\ \citenamefont
  {Pushkarev}}]{ZAKHAROV20041}%
  \BibitemOpen
  \bibfield  {author} {\bibinfo {author} {\bibfnamefont {V.}~\bibnamefont
  {Zakharov}}, \bibinfo {author} {\bibfnamefont {F.}~\bibnamefont {Dias}}, \
  and\ \bibinfo {author} {\bibfnamefont {A.}~\bibnamefont {Pushkarev}},\ }\href
  {\doibase 10.1016/j.physrep.2004.04.002} {\bibfield  {journal} {\bibinfo
  {journal} {Phys. Rep.}\ }\textbf {\bibinfo {volume} {398}},\ \bibinfo {pages}
  {1 } (\bibinfo {year} {2004})}\BibitemShut {NoStop}%
\bibitem [{\citenamefont {{Nazarenko}}(2011)}]{2011LNP825N}%
  \BibitemOpen
  \bibinfo {editor} {\bibfnamefont {S.}~\bibnamefont {{Nazarenko}}},\ ed.,\
  \href {\doibase 10.1007/978-3-642-15942-8} {\emph {\bibinfo {title} {Wave
  Turbulence, Lecture Notes in Physics}}},\ \bibinfo {series} {Berlin Springer
  Verlag}, Vol.\ \bibinfo {volume} {825}\ (\bibinfo {year} {2011})\BibitemShut
  {NoStop}%
\bibitem [{\citenamefont {Onorato}\ \emph {et~al.}(2015)\citenamefont
  {Onorato}, \citenamefont {Vozella}, \citenamefont {Proment},\ and\
  \citenamefont {Lvov}}]{Onorato4208}%
  \BibitemOpen
  \bibfield  {author} {\bibinfo {author} {\bibfnamefont {M.}~\bibnamefont
  {Onorato}}, \bibinfo {author} {\bibfnamefont {L.}~\bibnamefont {Vozella}},
  \bibinfo {author} {\bibfnamefont {D.}~\bibnamefont {Proment}}, \ and\
  \bibinfo {author} {\bibfnamefont {Y.~V.}\ \bibnamefont {Lvov}},\ }\href
  {\doibase 10.1073/pnas.1404397112} {\bibfield  {journal} {\bibinfo  {journal}
  {Proc. Natl. Acad. Sci. U.S.A.}\ }\textbf {\bibinfo {volume} {112}},\
  \bibinfo {pages} {4208} (\bibinfo {year} {2015})}\BibitemShut {NoStop}%
\bibitem [{\citenamefont {Lvov}\ and\ \citenamefont
  {Onorato}(2018)}]{PhysRevLett.120.144301}%
  \BibitemOpen
  \bibfield  {author} {\bibinfo {author} {\bibfnamefont {Y.~V.}\ \bibnamefont
  {Lvov}}\ and\ \bibinfo {author} {\bibfnamefont {M.}~\bibnamefont {Onorato}},\
  }\href {\doibase 10.1103/PhysRevLett.120.144301} {\bibfield  {journal}
  {\bibinfo  {journal} {Phys. Rev. Lett.}\ }\textbf {\bibinfo {volume} {120}},\
  \bibinfo {pages} {144301} (\bibinfo {year} {2018})}\BibitemShut {NoStop}%
\bibitem [{\citenamefont {Pistone}\ \emph
  {et~al.}(2018{\natexlab{a}})\citenamefont {Pistone}, \citenamefont
  {Onorato},\ and\ \citenamefont {Chibbaro}}]{0295-5075-121-4-44003}%
  \BibitemOpen
  \bibfield  {author} {\bibinfo {author} {\bibfnamefont {L.}~\bibnamefont
  {Pistone}}, \bibinfo {author} {\bibfnamefont {M.}~\bibnamefont {Onorato}}, \
  and\ \bibinfo {author} {\bibfnamefont {S.}~\bibnamefont {Chibbaro}},\ }\href
  {http://stacks.iop.org/0295-5075/121/i=4/a=44003} {\bibfield  {journal}
  {\bibinfo  {journal} {EPL (Europhysics Letters)}\ }\textbf {\bibinfo {volume}
  {121}},\ \bibinfo {pages} {44003} (\bibinfo {year}
  {2018}{\natexlab{a}})}\BibitemShut {NoStop}%
\bibitem [{\citenamefont {Fu}\ \emph {et~al.}(2018)\citenamefont {Fu},
  \citenamefont {Zhang},\ and\ \citenamefont {Zhao}}]{Our2018}%
  \BibitemOpen
  \bibfield  {author} {\bibinfo {author} {\bibfnamefont {W.}~\bibnamefont
  {Fu}}, \bibinfo {author} {\bibfnamefont {Y.}~\bibnamefont {Zhang}}, \ and\
  \bibinfo {author} {\bibfnamefont {H.}~\bibnamefont {Zhao}},\ }\href
  {https://arxiv.org/abs/1811.05697} {\bibfield  {journal} {\bibinfo  {journal}
  {arXiv:1811.05697}\ } (\bibinfo {year} {2018})}\BibitemShut {NoStop}%
\bibitem [{\citenamefont {Fu}\ \emph {et~al.}(2019)\citenamefont {Fu},
  \citenamefont {Zhang},\ and\ \citenamefont {Zhao}}]{Fu_2019}%
  \BibitemOpen
  \bibfield  {author} {\bibinfo {author} {\bibfnamefont {W.}~\bibnamefont
  {Fu}}, \bibinfo {author} {\bibfnamefont {Y.}~\bibnamefont {Zhang}}, \ and\
  \bibinfo {author} {\bibfnamefont {H.}~\bibnamefont {Zhao}},\ }\href {\doibase
  10.1088/1367-2630/ab115a} {\bibfield  {journal} {\bibinfo  {journal} {New J.
  Phys.}\ }\textbf {\bibinfo {volume} {21}},\ \bibinfo {pages} {043009}
  (\bibinfo {year} {2019})}\BibitemShut {NoStop}%
\bibitem [{\citenamefont {Pistone}\ \emph
  {et~al.}(2018{\natexlab{b}})\citenamefont {Pistone}, \citenamefont
  {Chibbaro}, \citenamefont {Bustamante}, \citenamefont {L'vov},\ and\
  \citenamefont {Onorato}}]{Pistone2018}%
  \BibitemOpen
  \bibfield  {author} {\bibinfo {author} {\bibfnamefont {L.}~\bibnamefont
  {Pistone}}, \bibinfo {author} {\bibfnamefont {S.}~\bibnamefont {Chibbaro}},
  \bibinfo {author} {\bibfnamefont {M.}~\bibnamefont {Bustamante}}, \bibinfo
  {author} {\bibfnamefont {Y.}~\bibnamefont {L'vov}}, \ and\ \bibinfo {author}
  {\bibfnamefont {M.}~\bibnamefont {Onorato}},\ }\href
  {https://arxiv.org/abs/1812.08279} {\bibfield  {journal} {\bibinfo  {journal}
  {arXiv:1812.08279}\ } (\bibinfo {year} {2018}{\natexlab{b}})}\BibitemShut
  {NoStop}%
\bibitem [{\citenamefont {Toda}(1989)}]{toda1989theory}%
  \BibitemOpen
  \bibfield  {author} {\bibinfo {author} {\bibfnamefont {M.}~\bibnamefont
  {Toda}},\ }\href {\doibase 10.1007/978-3-642-83219-2} {\emph {\bibinfo
  {title} {Theory of nonlinear lattices}}},\ \bibinfo {edition} {3rd}\ ed.,\
  Vol.~\bibinfo {volume} {20}\ (\bibinfo  {publisher} {Springer Cerkag Berlin
  Heidelberg},\ \bibinfo {year} {1989})\BibitemShut {NoStop}%
\bibitem [{\citenamefont {Casati}\ and\ \citenamefont
  {Ford}(1975)}]{PhysRevA.12.1702}%
  \BibitemOpen
  \bibfield  {author} {\bibinfo {author} {\bibfnamefont {G.}~\bibnamefont
  {Casati}}\ and\ \bibinfo {author} {\bibfnamefont {J.}~\bibnamefont {Ford}},\
  }\href {\doibase 10.1103/PhysRevA.12.1702} {\bibfield  {journal} {\bibinfo
  {journal} {Phys. Rev. A}\ }\textbf {\bibinfo {volume} {12}},\ \bibinfo
  {pages} {1702} (\bibinfo {year} {1975})}\BibitemShut {NoStop}%
\bibitem [{\citenamefont {Dash}\ and\ \citenamefont
  {Patnaik}(1981)}]{PhysRevA.23.959}%
  \BibitemOpen
  \bibfield  {author} {\bibinfo {author} {\bibfnamefont {P.~C.}\ \bibnamefont
  {Dash}}\ and\ \bibinfo {author} {\bibfnamefont {K.}~\bibnamefont {Patnaik}},\
  }\href {\doibase 10.1103/PhysRevA.23.959} {\bibfield  {journal} {\bibinfo
  {journal} {Phys. Rev. A}\ }\textbf {\bibinfo {volume} {23}},\ \bibinfo
  {pages} {959} (\bibinfo {year} {1981})}\BibitemShut {NoStop}%
\bibitem [{\citenamefont {Mokross}\ and\ \citenamefont
  {B\"uttner}(1981)}]{PhysRevA.24.2826}%
  \BibitemOpen
  \bibfield  {author} {\bibinfo {author} {\bibfnamefont {F.}~\bibnamefont
  {Mokross}}\ and\ \bibinfo {author} {\bibfnamefont {H.}~\bibnamefont
  {B\"uttner}},\ }\href {\doibase 10.1103/PhysRevA.24.2826} {\bibfield
  {journal} {\bibinfo  {journal} {Phys. Rev. A}\ }\textbf {\bibinfo {volume}
  {24}},\ \bibinfo {pages} {2826} (\bibinfo {year} {1981})}\BibitemShut
  {NoStop}%
\bibitem [{\citenamefont {Diederich}(1985)}]{Diederich_1985}%
  \BibitemOpen
  \bibfield  {author} {\bibinfo {author} {\bibfnamefont {S.}~\bibnamefont
  {Diederich}},\ }\href {\doibase 10.1088/0022-3719/18/18/008} {\bibfield
  {journal} {\bibinfo  {journal} {J. Phys. C: Solid State Phys.}\ }\textbf
  {\bibinfo {volume} {18}},\ \bibinfo {pages} {3415} (\bibinfo {year}
  {1985})}\BibitemShut {NoStop}%
\bibitem [{\citenamefont {Aoki}\ and\ \citenamefont
  {Takeno}(1995)}]{aoki1995stationary}%
  \BibitemOpen
  \bibfield  {author} {\bibinfo {author} {\bibfnamefont {M.}~\bibnamefont
  {Aoki}}\ and\ \bibinfo {author} {\bibfnamefont {S.}~\bibnamefont {Takeno}},\
  }\href {\doibase 10.1143/JPSJ.64.809} {\bibfield  {journal} {\bibinfo
  {journal} {J. Phys. Soc. Jpn.}\ }\textbf {\bibinfo {volume} {64}},\ \bibinfo
  {pages} {809} (\bibinfo {year} {1995})}\BibitemShut {NoStop}%
\bibitem [{\citenamefont {Mokross}\ and\ \citenamefont
  {Buttner}(1983)}]{Mokross_1983}%
  \BibitemOpen
  \bibfield  {author} {\bibinfo {author} {\bibfnamefont {F.}~\bibnamefont
  {Mokross}}\ and\ \bibinfo {author} {\bibfnamefont {H.}~\bibnamefont
  {Buttner}},\ }\href {\doibase 10.1088/0022-3719/16/23/015} {\bibfield
  {journal} {\bibinfo  {journal} {J. Phys. C: Solid State Phys.}\ }\textbf
  {\bibinfo {volume} {16}},\ \bibinfo {pages} {4539} (\bibinfo {year}
  {1983})}\BibitemShut {NoStop}%
\bibitem [{\citenamefont {Jackson}\ and\ \citenamefont
  {Mistriotis}(1989)}]{jackson1989thermal}%
  \BibitemOpen
  \bibfield  {author} {\bibinfo {author} {\bibfnamefont {E.~A.}\ \bibnamefont
  {Jackson}}\ and\ \bibinfo {author} {\bibfnamefont {A.~D.}\ \bibnamefont
  {Mistriotis}},\ }\href {\doibase 10.1088/0953-8984/1/7/006} {\bibfield
  {journal} {\bibinfo  {journal} {J. Phys.: Condens. Matter}\ }\textbf
  {\bibinfo {volume} {1}},\ \bibinfo {pages} {1223} (\bibinfo {year}
  {1989})}\BibitemShut {NoStop}%
\bibitem [{\citenamefont {Hatano}(1999)}]{PhysRevE.59.R1}%
  \BibitemOpen
  \bibfield  {author} {\bibinfo {author} {\bibfnamefont {T.}~\bibnamefont
  {Hatano}},\ }\href {\doibase 10.1103/PhysRevE.59.R1} {\bibfield  {journal}
  {\bibinfo  {journal} {Phys. Rev. E}\ }\textbf {\bibinfo {volume} {59}},\
  \bibinfo {pages} {R1} (\bibinfo {year} {1999})}\BibitemShut {NoStop}%
\bibitem [{\citenamefont {Chen}\ \emph {et~al.}(2014)\citenamefont {Chen},
  \citenamefont {Wang}, \citenamefont {Casati},\ and\ \citenamefont
  {Benenti}}]{PhysRevE.90.032134}%
  \BibitemOpen
  \bibfield  {author} {\bibinfo {author} {\bibfnamefont {S.}~\bibnamefont
  {Chen}}, \bibinfo {author} {\bibfnamefont {J.}~\bibnamefont {Wang}}, \bibinfo
  {author} {\bibfnamefont {G.}~\bibnamefont {Casati}}, \ and\ \bibinfo {author}
  {\bibfnamefont {G.}~\bibnamefont {Benenti}},\ }\href {\doibase
  10.1103/PhysRevE.90.032134} {\bibfield  {journal} {\bibinfo  {journal} {Phys.
  Rev. E}\ }\textbf {\bibinfo {volume} {90}},\ \bibinfo {pages} {032134}
  (\bibinfo {year} {2014})}\BibitemShut {NoStop}%
\bibitem [{\citenamefont {Chaturvedi}\ and\ \citenamefont
  {Baijal}(1974)}]{Diatomic1974}%
  \BibitemOpen
  \bibfield  {author} {\bibinfo {author} {\bibfnamefont {D.~K.}\ \bibnamefont
  {Chaturvedi}}\ and\ \bibinfo {author} {\bibfnamefont {J.~S.}\ \bibnamefont
  {Baijal}},\ }\href {\doibase 10.1119/1.1987756} {\bibfield  {journal}
  {\bibinfo  {journal} {Am. J. Phys}\ }\textbf {\bibinfo {volume} {42}},\
  \bibinfo {pages} {482} (\bibinfo {year} {1974})}\BibitemShut {NoStop}%
\bibitem [{\citenamefont {Leon}\ \emph {et~al.}(2013)\citenamefont {Leon},
  \citenamefont {Bj{\"o}rck},\ and\ \citenamefont {Gander}}]{leon2013gram}%
  \BibitemOpen
  \bibfield  {author} {\bibinfo {author} {\bibfnamefont {S.~J.}\ \bibnamefont
  {Leon}}, \bibinfo {author} {\bibfnamefont {{\AA}.}~\bibnamefont
  {Bj{\"o}rck}}, \ and\ \bibinfo {author} {\bibfnamefont {W.}~\bibnamefont
  {Gander}},\ }\href {\doibase 10.1002/nla.1839} {\bibfield  {journal}
  {\bibinfo  {journal} {Numerical Linear Algebra with Applications}\ }\textbf
  {\bibinfo {volume} {20}},\ \bibinfo {pages} {492} (\bibinfo {year}
  {2013})}\BibitemShut {NoStop}%
\bibitem [{\citenamefont {Livi}\ \emph {et~al.}(1985)\citenamefont {Livi},
  \citenamefont {Pettini}, \citenamefont {Ruffo}, \citenamefont
  {Sparpaglione},\ and\ \citenamefont {Vulpiani}}]{PhysRevA.31.1039}%
  \BibitemOpen
  \bibfield  {author} {\bibinfo {author} {\bibfnamefont {R.}~\bibnamefont
  {Livi}}, \bibinfo {author} {\bibfnamefont {M.}~\bibnamefont {Pettini}},
  \bibinfo {author} {\bibfnamefont {S.}~\bibnamefont {Ruffo}}, \bibinfo
  {author} {\bibfnamefont {M.}~\bibnamefont {Sparpaglione}}, \ and\ \bibinfo
  {author} {\bibfnamefont {A.}~\bibnamefont {Vulpiani}},\ }\href {\doibase
  10.1103/PhysRevA.31.1039} {\bibfield  {journal} {\bibinfo  {journal} {Phys.
  Rev. A}\ }\textbf {\bibinfo {volume} {31}},\ \bibinfo {pages} {1039}
  (\bibinfo {year} {1985})}\BibitemShut {NoStop}%
\bibitem [{\citenamefont {Goedde}\ \emph {et~al.}(1992)\citenamefont {Goedde},
  \citenamefont {Lichtenberg},\ and\ \citenamefont
  {Lieberman}}]{GOEDDE1992200}%
  \BibitemOpen
  \bibfield  {author} {\bibinfo {author} {\bibfnamefont {C.}~\bibnamefont
  {Goedde}}, \bibinfo {author} {\bibfnamefont {A.}~\bibnamefont {Lichtenberg}},
  \ and\ \bibinfo {author} {\bibfnamefont {M.}~\bibnamefont {Lieberman}},\
  }\href {\doibase 10.1016/0167-2789(92)90216-A} {\bibfield  {journal}
  {\bibinfo  {journal} {Physica D}\ }\textbf {\bibinfo {volume} {59}},\
  \bibinfo {pages} {200 } (\bibinfo {year} {1992})}\BibitemShut {NoStop}%
\bibitem [{\citenamefont {Yoshida}(1990)}]{YOSHIDA1990262}%
  \BibitemOpen
  \bibfield  {author} {\bibinfo {author} {\bibfnamefont {H.}~\bibnamefont
  {Yoshida}},\ }\href {\doibase 10.1016/0375-9601(90)90092-3} {\bibfield
  {journal} {\bibinfo  {journal} {Phys. Lett. A}\ }\textbf {\bibinfo {volume}
  {150}},\ \bibinfo {pages} {262 } (\bibinfo {year} {1990})}\BibitemShut
  {NoStop}%
\end{thebibliography}%

\end{document}